\documentclass{epl} 
\newcommand{\br}{{\bf r}}
\newcommand{\bq}{{\bf q}}

\title{Finite-size effects on the Hamiltonian dynamics \\ of the XY-model}
\shorttitle{Finite-size effects in the XY-model}
\author{S. Lepri \inst{1,2}\thanks{E-mail: {\tt lepri@avanzi.de.unifi.it}}, 
S. Ruffo \inst{1,2}\thanks{E-mail: {\tt ruffo@avanzi.de.unifi.it} 
also at INFN, Firenze}}
\institute{
\inst{1} Dipartimento di Energetica ``S. Stecco'',
Via Santa Marta 3, I-50139 Florence, Italy \\
\inst{2} Istituto Nazionale di Fisica della Materia, Unit\`a di Firenze, 
L.go E. Fermi 3 I-50125 Florence, Italy
}

\date{\today}
\begin{document}
\maketitle 
\begin{abstract}
The dynamical properties of the finite-size magnetization $M$ in the critical 
region $T \leq T_{KTB}$ of the planar rotor model on a $L \times L$ square
lattice are analyzed by means of microcanonical simulations . The behavior 
of the ${\bf q}=0$ structure factor
at high frequencies is consistent with field-theoretical results, but
new additional features occur at lower frequencies.
The motion of $M$ determines a region of spectral lines and 
the presence of a central peak, which we attribute to
phase diffusion. 
Near $T_{KTB}$ the diffusion constant scales with system size as 
$D \sim L^{-1.6(3)}$.
\vspace{0.2cm}
\end{abstract}
\vspace{0.2cm}

The XY or ``planar rotor" model consists of a set of classical spins 
${\bf S}_\br$ of unit length confined in a plane, whose orientation 
is specified by the angle $\theta_\br$, $\br$ being the position 
vector on a square lattice of size $N=L \times L$ with periodic boundary 
conditions. It is known that this model does not admit
equations of motion and therefore its {\it canonical} dynamics is usually 
simulated either by Monte-Carlo methods~\cite{tobo,gupta1} or by Langevin type
equations~\cite{loft,jensen}. {\it Microcanonical} approaches consist
instead, either in considering a three component spin model~\cite{evertz} or
into adding a kinetic energy term~\cite{kogut}. The latter method
can be also generally applied to other physical systems (see e.g. the
application to lattice gauge theories~\cite{calla}). All these methods
should display the same static properties, as it has been 
verified up to some extent~\cite{kogut,leoncini,bram}. 

One of the most striking features of the XY model is the presence of 
strong finite-size effects~\cite{bramwell}, e.g. the existence of a
sizeable magnetization for large samples, despite the fact that
long-range order cannot occur for the infinite system. For this reason, we 
are motivated to study the dynamical features of a finite lattice, a 
topic which has been far less investigated. In this respect the choice
of the dynamics, e.g. canonical vs. microcanonical, is expected
to play a crucial role.

In this Letter we consider the following Hamiltonian
\begin{equation}
{\cal H} \;= \;\sum_\br\frac{p_\br^2}{2}+ \sum_{\langle \br, \br' \rangle} 
[1-\cos(\theta_{\br'}-\theta_\br)]\;,
\label{hami}
\end{equation}
where $p_\br=\dot{\theta_\br}$ is the angular momentum of the rotor and 
the sum ranges over the four nearest neighbors of site $\br$~\cite{nota0}.
We have set both the inertia of the rotors and the ferromagnetic 
coupling constant to unity so that the only physical control parameter is 
the energy per spin $e={\cal H}/N$. Actually, there is a second constant 
of the motion, the total angular momentum  $P = \sum_\br p_\br$, whose
choice affects the results in a trivial way. In the numerical simulations
we set $P=0$ to avoid global ballistic rotation. 

A previous study of the static properties of (\ref{hami})~\cite{leoncini} 
showed that the system undergoes a Kosterlitz-Thouless-Berezinskii 
(KTB) transition~\cite{gulacsi} at $e=e_{KTB}\approx 1.0$ corresponding 
to a kinetic temperature $T_{KTB}\approx 0.89$, which is in agreement with
the value 0.894(5) obtained in the canonical ensemble~\cite{gupta1}.

Here, we focus our attention on the statistical behaviour of the 
istantaneous complex magnetization $M=M_x+i M_y$, defined as
\begin{equation}
M \;=\;\frac{1}{L^2}\sum_{\br} \, {\bf S}_\br  
\;=\;\frac{1}{L^2}\sum_{\br} \, e^{i\theta_\br} \;=\; m\,e^{i\psi}~,
\label{magne}
\end{equation}
in the ordered phase $e < e_{KTB}$. In this regime the phase $\psi$ is
well defined, because $m$ is bounded away from zero~\cite{archa}.
Moreover, $\langle m\rangle$ is a sizeable quantity and can be regarded 
as a pseudo ({\it i.e.} non-extensive) order parameter
($\langle \cdot \rangle$ denotes the ensemble average). Indeed, it vanishes 
only algebraically with system size, $\langle m \rangle= (2N)^{-k_BT/8 \pi}$ 
in the spin-wave limit~\cite{bramwell} .

The numerical integration of the equations of motion is performed using the
fourth-order McLahlan-Atela algorithm~\cite{atela} which is an explicit 
scheme constructed from a suitable truncation of the evolution operator that
preserves the Hamiltonian structure. One of the major merits of symplectic
algorithms is that the error on the energy does not increase with the length of
the run. The chosen time step (0.01-0.05 in our units) ensures that in every
simulation energy fluctuates around the prescribed value with a relative
accuracy below $10^{-5}$.

The dynamics is fully characterized by the structure factor 
$S({\bf q},\omega)$, {\it i.e.} the space-time Fourier transform
of the the spin-spin correlation function (we assume translational invariance) 
\begin{equation}
S(\br,t) \;=\; \langle e^{i[\theta_\br (t)-\theta_{\bf 0}(0)]}\rangle
\quad .
\label{correla}
\end{equation}
Accordingly, the power spectrum of $M$ is proportional to the zero 
wavenumber $({\bf q}=0)$ 
structure factor
\begin{equation}
S({\bf 0},\omega) \;\propto\; \int_{-\infty}^{\infty} 
\langle M(t)M^*(0) \rangle e^{i\omega t} \, dt \;=\; 
\langle|M(\omega)|^2\rangle~.
\end{equation}

\begin{figure} 
%\centering\epsfig{figure=spettro1.eps,width=10cm,angle=-90}
%\centering\epsfig{figure=spettro2.eps,width=10cm,angle=-90}
\twofigures[scale=0.33,angle=-90]{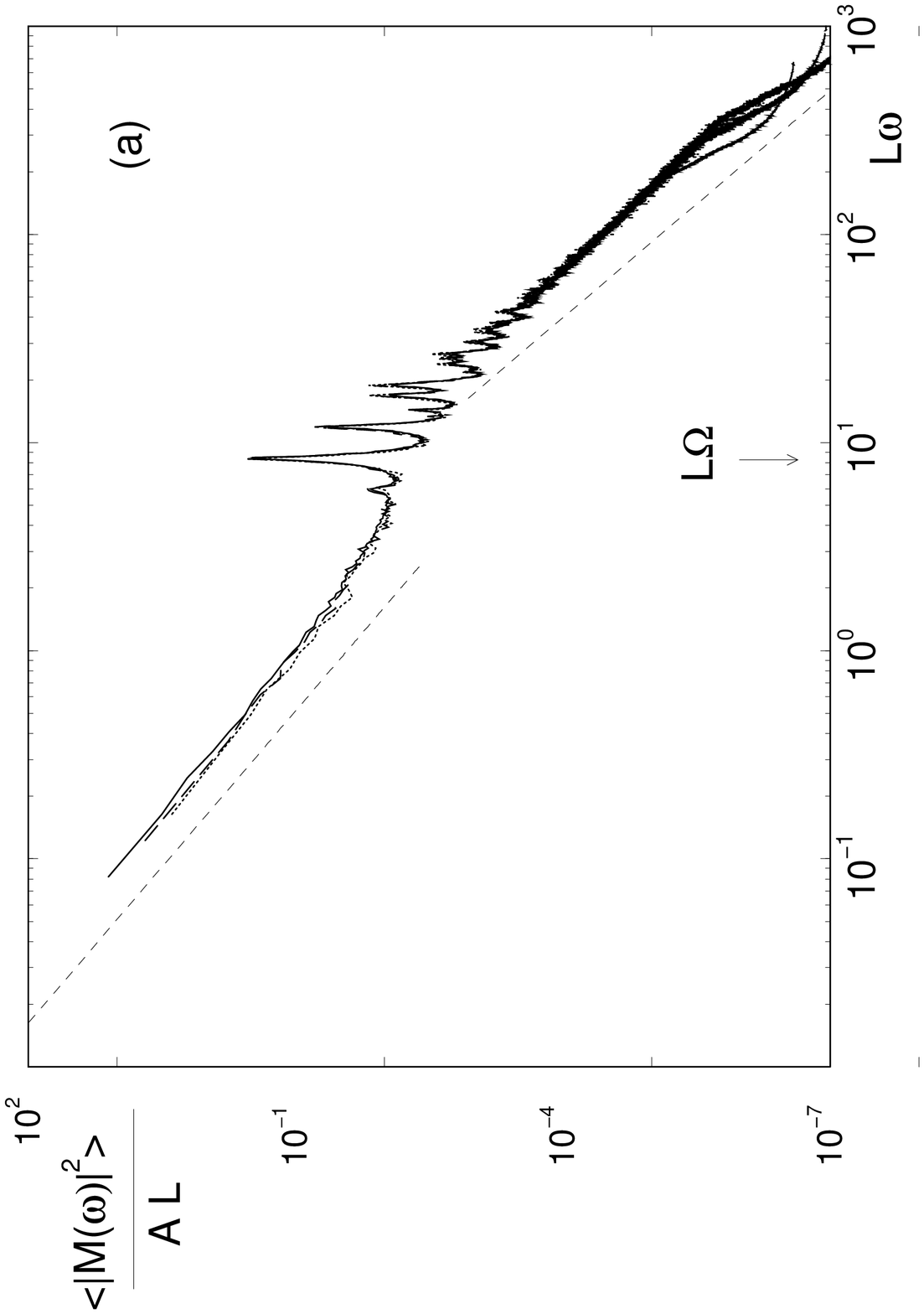}{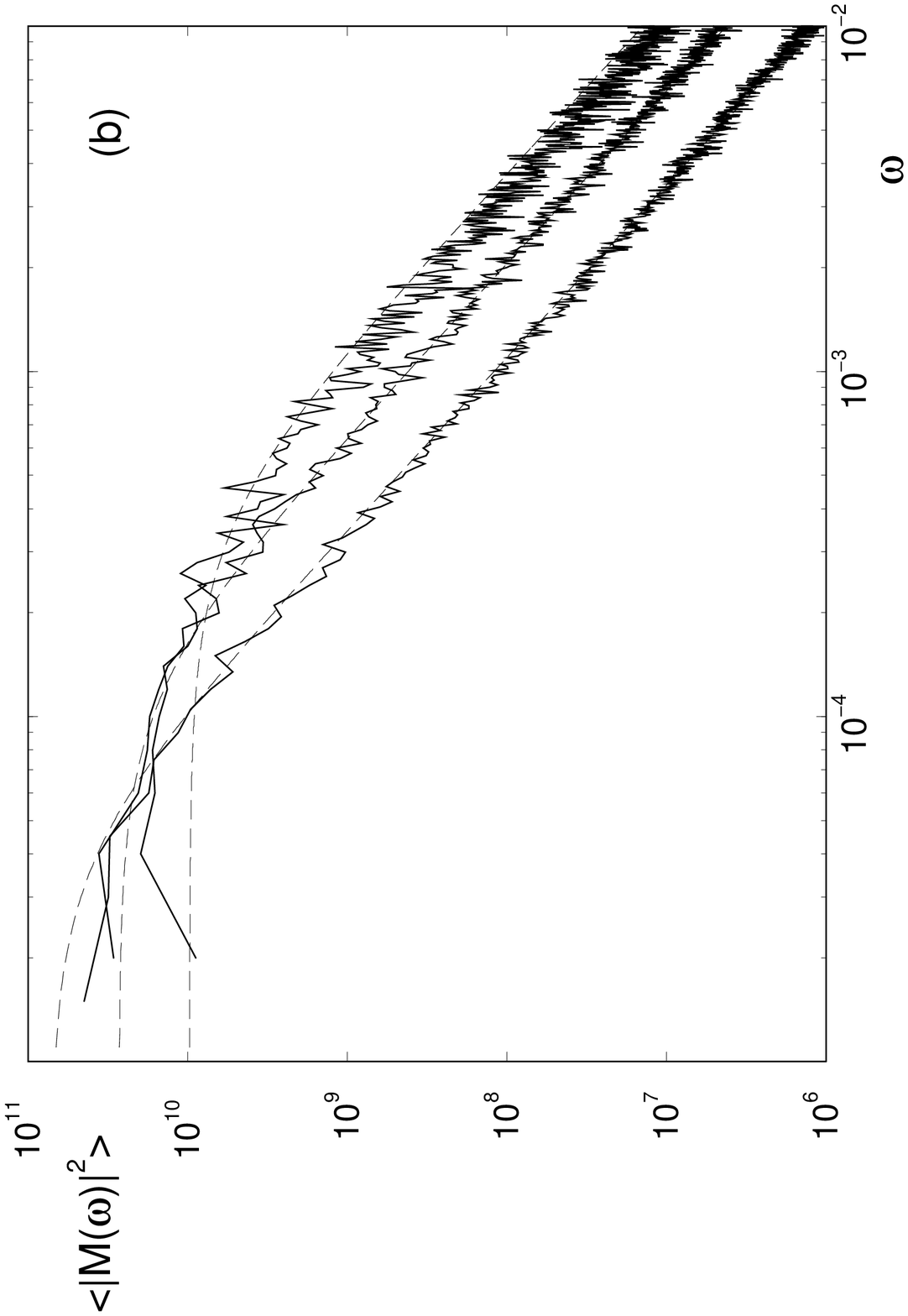} 
\caption{(a) Scaled high-frequency part of 
the spectrum $\langle|M(\omega)|^2\rangle$ for $e=0.992$ and 
$L=64,96,128$.  Each curve is an average of $\sim 10^3$ 
runs, each of $5\, 10^3$ time units (sampling time 0.3).
The thin dashed lines represent the power-laws $\omega^{-2}$ 
due to diffusion and the NF decay $\omega^{-2.75}$. 
(b) Low-frequency part of $\langle|M(\omega)|^2\rangle$ obtained from 
longer runs ($3 \, 10^5$ time units) for $L=8 ,16 , 32$ (from top to bottom).
The spectra are not scaled and are in arbitrary units.
Thin dashed lines are Lorentzian fits $C/(D^2+\omega^2)$.
}
\label{spettri}
\end{figure}

The dynamical finite-size scaling hypothesis suggests that this quantity should be
written as $\langle|M(\omega)|^2\rangle/AL^z=f(\omega L)$, where $A$ is the area
below the spectrum~\cite{evertz}. According to the theory by Nelson and Fisher
(NF)~\cite{nelson} we expect $z=1$ in our case. Our numerical results are in good
agreement with these predictions, see Fig.~\ref{spettri}a. Moreover, the
high-frequency part ($L\omega\gg 1$)  displays a power-law decay over
approximatively one decade. This is a prediction of NF theory which, relying on a
hydrodynamic description,  is expected to hold in this frequency range,
corresponding to times much shorter than the typical transit time  of a wave on
the lattice.    Obviously  a cutoff  frequency of order unity is imposed by the 
lattice discreteness. Although an accurate estimation of the decay exponent is
difficult in such a narrow range, the available data are consistent with the
predicted law $\omega^{-3+\eta}$ (with $\eta=k_BT/2\pi$)~\cite{note}, at least up
to the energy where a significant density of vortex pairs appears. This is
precisely the case at $e=0.992$ where we found $\omega^{-2.5(6)}$ which is
actually significantly different from the NF value $3-\eta$ with
$\eta\approx\eta(T_{KTB})=1/4$.  Notice that this result is somehow in contrast
with the theoretical expectations~\cite{gulacsi} as it corresponds to $\eta>1/4$.
Similar deviations are indeed observed in the static case and are
traced back to the presence of multiplicative logarithmic corrections.
Unfortunately, even the most accurate numerical result are controversial and
apparently in contrast with the theory (see \cite{janke} and references within
for a thoughtful account on this  issue). We thus limit ourselves to observe
that, at variance with what is reported in the literature we find $\eta>1/4$ in the
critical region.

The additional structure present in the spectrum at lower frequencies is beyond 
the validity range of NF theory and is related to the long-time dynamics of $M$
for the finite lattice. Indeed, it has been previously observed, in both
Monte-Carlo~\cite{archa} and microcanonical~\cite{clementi} simulations, that $M$ fluctuates
in time in the complex plane in a narrow region around the circle of radius
$\langle m\rangle$ centered in the origin. A closer inspection to our data,
reveals that such a motion has two main components. On one hand, $m(t)$
oscillates irregularly around its  mean value, with a typical frequency $\Omega$
which is inversely  proportional to $L$. On the other
hand, the phase $\psi(t)$  performs a slower random motion whose displacement
rapidly decreases  with the energy. For instance, already below $e \simeq 0.3$
the phase changes  of $M$  are hardly observable on the typical size and time
scales  of our simulations ($10^3-10^4$ time units).

These features of the motion of $M$ reflect into some distinctive properties of
$S({\bf 0},\omega)$. More precisely, the oscillations of $m$ manifest themselves
as a very  sharp line at $\Omega$ (Fig.~\ref{spettri}a), whereas the slower phase
motion determines a low frequency  ($\omega\ll\Omega$) component centered at
$\omega=0$~(Fig.~\ref{spettri}b). The latter is naturally interpreted as a
signature of phase diffusion, once  we assume $\psi$ to be a Brownian variable
with a diffusion time  $D^{-1}$ much larger than the typical time scale of the
motion of $m$. Accordingly, the spectrum is well fitted by a Lorentzian curve 
$C/(D^2+\omega^2)$, which allows also to determine $D$. 
Furthermore, the relative weight of the diffusive component of 
$\langle|M(\omega)|^2\rangle$ decreases upon decreasing $e$, in agreement 
with the qualitative observation that phase motion becomes less 
and less pronounced.

\begin{figure} 
%\centering\epsfig{figure=phdiff.eps,width=12cm,angle=-90}
\twofigures[scale=0.33,angle=-90]{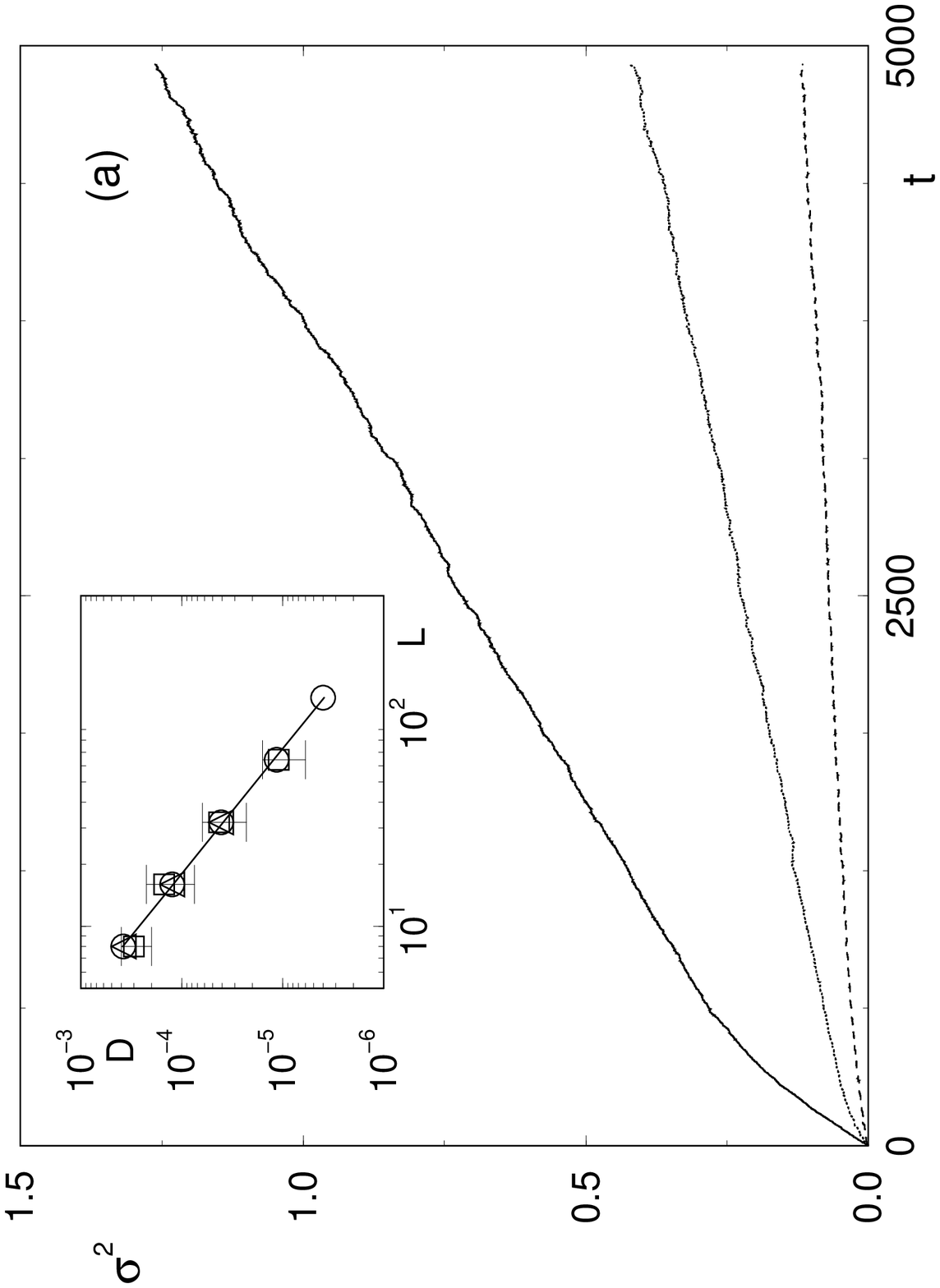}{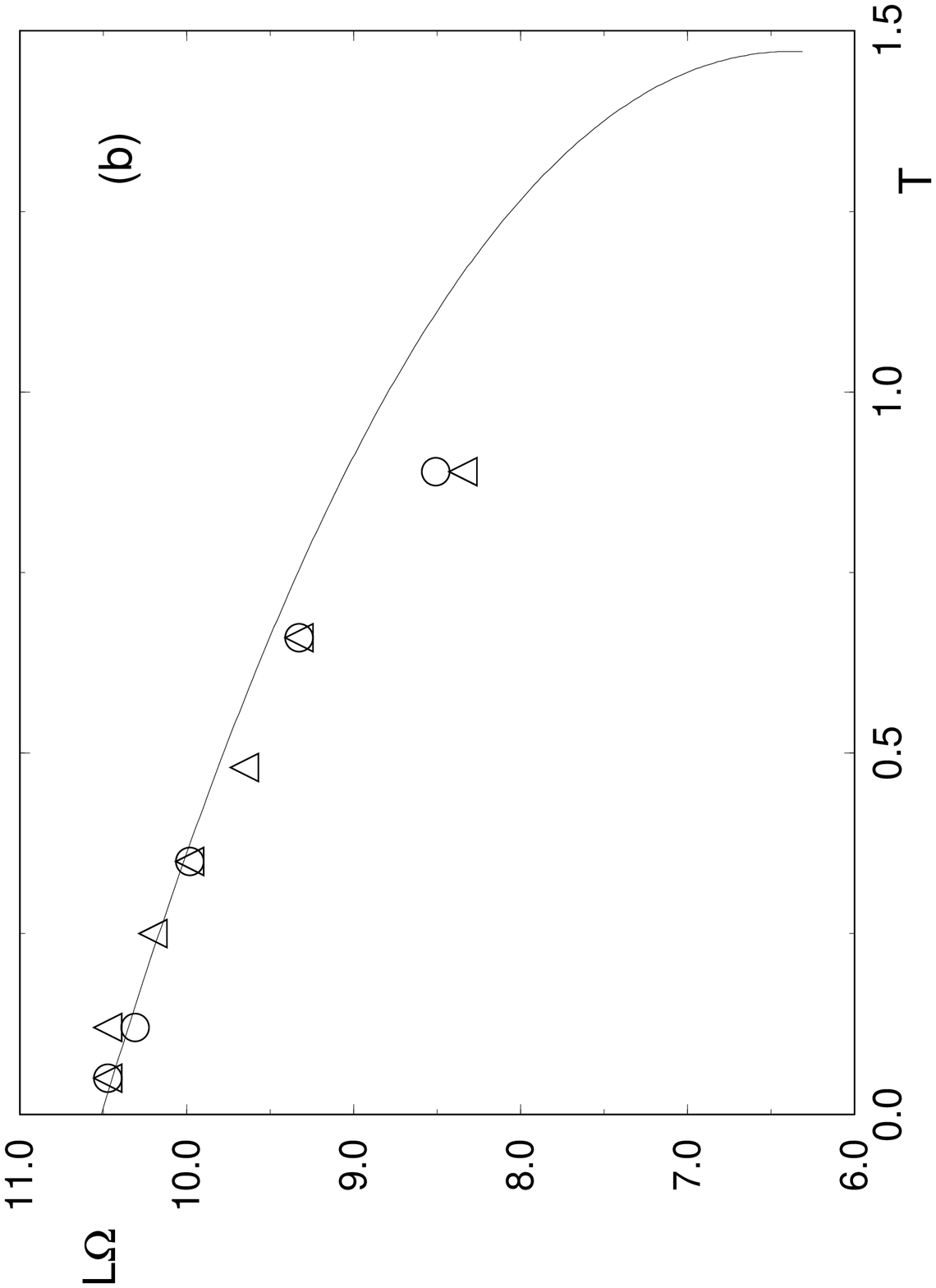}
\caption{(a) Mean squared displacement $\sigma^2=\langle(\psi-\psi_0)^2\rangle$
at $e=0.992$  for $L=16, 32, 64$ (from top to 
bottom). The inset shows the corresponding diffusion coefficient
$D$ evaluated from a linear fit (circles), from Green-Kubo
formula (squares) and from a Lorentzian fit of the low-frequency
region of $\langle|M(\omega)|^2\rangle$ (see Fig. \ref{spettri}b).
(b) Scaled oscillation frequency $L\Omega$ versus
the kinetic temperature for $L=64, 128$ (circles and triangles 
respectively). The solid line is proportional to $c(T)$ (see text).
}
\label{diff}
\end{figure}

To further support this interpretation, we have computed the time evolution of
the mean squared value $\sigma^2=\langle(\psi-\psi_0)^2\rangle$ for several 
lattice sizes. As shown in Fig.~\ref{diff}a, the data are consistent with
a diffusive motion, {\it i.e.} $\sigma^2=2Dt$, for large enough $t$. 
We have also performed an independent measure of the diffusion
constant based on Green-Kubo formula 
\begin{equation}
D \;=\; {1\over k_BT}\,\int_0^\infty \, \langle\dot\psi(t) \dot\psi(0)\rangle 
\,dt \quad .
\end{equation}
Remarkably, the values computed using the three methods agree extremely well 
within the statistical accuracy (better than 15$\%$ in the worst case - 
see the inset of Fig.~\ref{diff}a).

In analogy with what happens for the static magnetization  we expect the
diffusion coefficient to approach zero on increasing the lattice size. 
Indeed, we found evidence that $D$ vanishes algebraically  with a nontrivial exponent: a
best fit to the data in the inset of  Fig.~\ref{diff} gives $D \propto
L^{-1.6(3)}$ in the  range $8\le L \le 128$. Although we did not attempt 
to perform a
systematic study, we have observed that a similar scaling holds also at lower
energies. For instance, at $e=0.70$ we found a  sligthly smaller exponent
$\approx 1.5$ in the same range of sizes. Thus, we cannot rule out the
possibility that such exponent depends on temperature in analogy with the one of
finite-size magnetization.

The oscillations of $m$ are related to the
spin-wave modes with the longest wavelength. We thus expect $\Omega$ 
to be proportional to $c/L$, where $c$ is the renormalized spin-wave 
velocity appearing in the effective dispersion relation 
\begin{equation}
\omega_{\bf q}^2 \;=\; 4c^2 \,\left[\sin^2{q_x\over2} +  \sin^2{q_y\over2}
\right]~.
\label{omegaq}
\end{equation}
Neglecting the contribution of vortices, $c$ is only 
determined by the anharmonic interaction among the spin-wave modes,
thus leading to the estimate $-8c^2\log c = k_BT$ \cite{leoncini}. 
The data reported in Fig.~\ref{diff}b
show that this results accurately describes the dependence of $\Omega$ on  $T$
at least for low enough temperatures. Systematic  deviation at larger
temperature have of course to be expected due to the appearance  of
vortices. Indeed, it is usually believed~\cite{cote} that their effect 
is to further diminish the spin-wave velocity, which is in qualitative 
agreement with our data. 

\begin{figure} 
%\centering\epsfig{figure=SRT-rescale.eps,width=12cm,angle=-90}
\onefigure[scale=0.33,angle=-90]{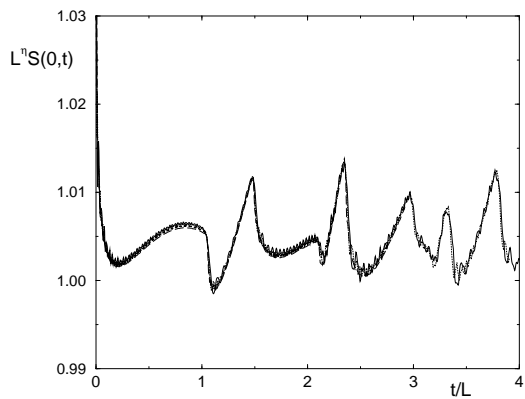} 
\caption{Finite-size scaling of the correlation function
$S(0,t)$, formula (\protect\ref{pertu}) for $k_BT=0.05$ and 
$L=128, 256, 512, 1024$. The scaling exponent is 
$\eta=k_BT/2\pi \approx 0.0079\dots$.
}
\label{srt}

\end{figure}

In order to better illustrate the new effects introduced by the finite 
lattice, we have evaluated the correlation (\ref{correla}) in the 
low-temperature (spin-wave) 
region to lowest order (Gaussian theory). The resulting expression
\begin{equation}
S(\br,t)\;=\;\exp \left[ {k_BT\over L^2} \, \sum_{\bq\ne {\bf 0}}
{\cos \omega_{\bf q}t \; e^{i  \bq \cdot \br } \,- \,1 \over
\omega_\bq ^2 }\right]
\label{pertu}
\end{equation}
is the lattice version of the formula obtained by NF in the 
continuum~\cite{nelson}. Their results
can by summarized by saying  that for  $ct\gg |\br|$, $S(\br,t)$ should be
independent of $\br$ and vanish algebraically in time as $ (ct)^{-\eta}$.  We
have evaluated numerically formula (\ref{pertu})  by performing the sum over
the discrete set of allowed ${\bf q}$s for several values of $r$. As a
first check, we have found that $S(\br,t)$ is practically independent of $\br$
for $r \ll L$. Therefore, we show in Fig.~\ref{srt} the behavior of $S(0,t)$ for
several values of $L$. We observe a dramatic difference with what expected from NF
theory: (i) the power law decay is restricted to a small region of short times
($t \ll L$) and rather well-defined oscillations appear with typical period of
order $L$ at larger times; (ii) the correlation function does not vanish
asymptotically,  as it should be for a system with nonzero magnetization. 
The oscillating behavior of (\ref{pertu}) is clearly due to the long-wavelength 
spin-waves and is therefore connected with the line structure 
around $\Omega$ in Fig.~\ref{spettri}a.
Hence, this signals again that beyond the limit of validity of NF theory new
effects due to the finite-size appear.

To summarize, in this Letter we have studied finite-size effects on the Hamiltonian
dynamics of the magnetization vector $M$ in the 2D XY model. We have
investigated the consequences of the motion of the modulus and the phase of $M$ on
the behavior of the ${\bf q}=0$ structure factor. They amount to the appearance of a diffusive
central component, which we associate to phase motion, and of a peak due to the
oscillations of the modulus. The high frequency part is basically consistent with
Nelson-Fisher theory~\cite{nelson}.  Similar properties are observed for other
types of dynamics~\cite{evertz}, for which the shape of  $S({\bf q},t)$ deviates from the one
given by the Nelson-Fisher  theory, exactly for the presence of a central peak. We
conjecture that such features might be related to our results, and a more
systematic comparison would be therefore desirable.  

Furthermore, we have found a convincing evidence that the diffusion constant $D$
vanishes algebraically with $L$ with an exponent different from $z=1$. If this
would be confirmed for larger lattices, one should conclude that the correlation
function does not exhibit a single scaling with $L^z$, which in turn would imply
a violation of the dynamic scaling hypothesis. This should be signalled
by a bad superposition of the structure factor below $\omega\sim D$ for 
different sizes. A direct test of this effect would however require simulations 
on much longer times.
Since scaling violations indeed occur for coarsening phenomena~\cite{bray},
which are far off-equilibrium processes, it would interesting to investigate
further this possibility also close to equilibrium. 

\acknowledgments
 
We acknowledge useful discussions with H. Chat\'e, C. Godr\`eche,  P. Simon.
We thank  P. C. W. Holdsworth for a careful reading of the  manuscript. 
This work is supported by the INFM-PAIS project {\it
Equilibrium and nonequilibrium dynamics in condensed matter}.

%\end{references}

%\newpage 


\begin{thebibliography}{0} 
%\begin{references}

\bibitem{tobo} 
\Name{J. Tobochnik \and G.V. Chester} 
\REVIEW{Phys. Rev. B}{20}{1979}{3761}.

\bibitem{gupta1} 
\Name{R. Gupta \and C.F. Baille} 
\REVIEW{Phys. Rev. B}{45}{1992}{2883}

\bibitem{loft} 
\Name{R. Loft \and T.A. DeGrand} 
\REVIEW{Phys. Rev. B}{35}{1987}{8528}.

\bibitem{jensen} 
\Name{L.M. Jensen, B.J. Kim \and P. Minnhagen} 
\REVIEW{Phys. Rev. B}{61}{2000}{15412}.

\bibitem{evertz} 
\Name{H.G. Evertz \and D.P. Landau} 
\REVIEW{Phys. Rev. B}{54}{1996}{12302}.
 
\bibitem{kogut} 
\Name{J. Kogut \and J. Polonyi} 
\REVIEW{Nucl. Phys. B}{[FS15]265}{1986}{313}.

\bibitem{calla}
\Name{D.J.E. Callaway \and A. Rahman} 
\REVIEW{Phys. Rev. Lett.}{49}{1982}{613}.

\bibitem{leoncini} 
\Name{X. Leoncini, A.D. Verga \and S. Ruffo} 
\REVIEW{Phys. Rev. E} {57}{1998}{6337}.

\bibitem{bram} 
\Name{S.T. Bramwell et al.}
\REVIEW{Phys. Rev. E} {63}{2001}{041106}.
 
\bibitem{bramwell} 
\Name{S.T. Bramwell \and P.C.W. Holdsworth}
\REVIEW{J. Phys. Condens. Matter} {5}{1993}{L53}.  

\bibitem{nota0} 
It could be shown that (\ref{hami}) is obtained as the classical limit of
a quantum Heisenberg Hamiltonian with an anisotropy term
$\sum_{\br} (S_{\br}^z)^2$, using the representation introduced
in \Name{J. Villain} \REVIEW{J. Phys. (Paris)}{35}{1974}{27}.

\bibitem{gulacsi} 
See \Name{Z. Gul\'acsi \and M. Gul\'acsi} 
\REVIEW{Adv. Phys.}{47}{1998} 1 for a comprehensive review.

\bibitem{archa} 
\Name{P. Archambault, S.T. Bramwell \and P.C.W. Holdsworth} 
\REVIEW{J. Phys. A }{30} {1997} {8363}. 

   
\bibitem{atela} 
\Name{P.I. McLachlan \and P. Atela} 
\REVIEW{Nonlinearity}{ 5}{1992}{541}.
 
\bibitem{nelson} 
\Name{D.R. Nelson \and D.S. Fisher}
\REVIEW{Phys. Rev. B}{16} {1977} {4945}.

\bibitem{note} For example, we checked that the data are 
compatible with the values 2.99 and 2.89 at $e=0.05$ and
0.70 respectively. Notice that here the nonlinear corrections 
are basically irrelevant: the effective value of 
$\eta=k_BT/2\pi J_{eff}$  alters the 
value of the bare exponent by less than 1\% in the latter case.


%\bibitem{gupta}
%Similar deviations were also observed in  
%\Name{R. Gupta et al.} 
%\REVIEW{Phys. Rev. Lett.}{61} {1988}{1996}.

\bibitem{janke} 
\Name{W. Janke}
\REVIEW{Phys. Rev. B}{55} {1997} {3580}.
 
\bibitem{clementi} 
\Name{M. Cerruti-Sola, C. Clementi \and M. Pettini} 
\REVIEW{Phys. Rev. E} {61} {2000} {5171}.
  

\bibitem{cote} 
\Name{R. Cot\`e \and A. Griffin} 
\REVIEW{Phys. Rev. B }{ 34}{1986}{6240}.

\bibitem{bray} 
\Name{A.J. Bray, A.J. Briant \and  D.K. Jervis} 
\REVIEW{Phys. Rev. Lett.}{84} {2000}{1503}.
 

\end{thebibliography}
\end{document}